 \documentclass[aip,jmp,amsmath,amssymb,reprint,floatfix,onecolumn]{revtex4}
\usepackage{bm} 
\usepackage{amssymb} 
\usepackage{natbib} 
\usepackage{stmaryrd}
\usepackage{graphicx} 
\usepackage{color}
\usepackage{amsmath} 
\usepackage{amsfonts}  
\usepackage{bm}
\usepackage{placeins}
\newcommand{\be}{\begin{equation}}
\newcommand{\ee}{\end{equation}}

\usepackage{color}

\usepackage{amsmath}
\usepackage{graphicx}
\newcommand{\ve}[1]{\ensuremath{\mbox{\boldmath$#1$}}}
\newcommand{\sve}[1]{\ensuremath{\mbox{\footnotesize\boldmath$#1$}}}

\newcommand{\T}{^{\rm T}}
\newcommand{\D}{{\rm D}}

\newcommand\nn{\nonumber}

\newcommand{\dvec}[1]{\ensuremath{\ve v_{\rm s}}}

\newcommand{\eqnlab}[1]{\label{eq:#1}}
\newcommand{\figlab}[1]{\label{fig:#1}}
\newcommand{\eqnref}[1]{(\ref{eq:#1})}

\newcommand{\Eqnref}[1]{Eq.~(\ref{eq:#1})}
\newcommand{\Figref}[1]{Fig.~\ref{fig:#1}}
\newcommand{\Secref}[1]{Section~\ref{sec:#1}}

\newcommand{\taus}{\tau_{\rm s}}
\newcommand{\tauK}{\tau_{\rm K}}
\DeclareMathOperator{\E}{{\mathcal F}}
\DeclareMathOperator{\erfc}{erfc}

\DeclareMathOperator{\ku}{Ku}
\DeclareMathOperator{\st}{St}
\DeclareMathOperator{\re}{Re}

\newcommand{\marker}[1]{\protect\raisebox{-0.5mm}{\protect\includegraphics[width=2.5mm,clip]{markBW#1.eps}}}

\begin{document}
 
\title{Inertial-particle accelerations in turbulence: a Lagrangian closure \footnote{postprint version of the article
 published on Journal Fluid Mech. vol. 798  pp. 187-200 (2016) }}
 
%\author[L. Biferale, C. Meneveau and R. Verzicco]{ L.\ns B\ls I\ls F\ls E\ls R\ls A\ls L\ls E$^{1}$, C. \ns M\ls E\ls N\ls E\ls V\ls E\ls A\ls U$^{2}$ and R.\ns V\ls E\ls R\ls Z\ls I\ls C\ls C\ls O $^{3}$}

\author{S. Vajedi$^1$, K. Gustavsson $^2$, B. Mehlig $^1$, and L. Biferale $^2$}

\affiliation{$^1$ Department of Physics, Gothenburg University,
 SE-41296 Gothenburg, Sweden \\
$^2$ Department of Physics and INFN, University of Rome 'Tor Vergata', Via della Ricerca Scientifica 1, 00133 Rome, Italy}

\begin{abstract}
The distribution of particle accelerations in turbulence is intermittent, with non-Gaussian tails
that are quite different for light and heavy particles.
In this article we analyse a closure scheme for the acceleration fluctuations of light and heavy
inertial particles in turbulence,
formulated in terms of Lagrangian correlation functions of fluid tracers.
We compute the variance and the flatness of inertial particle accelerations and we discuss their dependency on the  Stokes number.
The closure incorporates  effects induced by the Lagrangian correlations along the trajectories of fluid tracers, and its predictions agree well with results of direct numerical simulations of inertial particles in turbulence, provided that the effects induced by  the {\it inertial preferential sampling} of heavy/light particles outside/inside vortices are negligible. In particular, the scheme predicts the correct functional behaviour of the acceleration variance, as a function of $\st$, as well as
the presence of a minimum/maximum for the flatness of the acceleration of heavy/light  particles,
 in good qualitative agreement with numerical data.
We also show that the closure works  well when applied to the Lagrangian evolution of particles  using a stochastic surrogate for the underlying Eulerian velocity field.
Our results support the conclusion that  there exist important contributions to the statistics of the acceleration of inertial particles  independent of the preferential sampling.  For heavy particles we observe deviations between the  predictions of the
closure scheme and direct numerical simulations, at  Stokes numbers of order unity.
For light particles the deviation occurs for larger Stokes numbers.
%Finally, we discuss possible further improvement of the closure which should be able to capture corrections due to inertial preferential sampling.
%The Lagrangian closure scheme can be rigorously justified in perturbation theory of a statistical model for particles in turbulence. This model possesses a small parameter (the {\it Kubo number}) controlling persistence of flow structures.  The statistical model
%yields predictions that are in good qualitative agreement with the
% direct numerical simulations, but effects due to inertial preferential sampling are much weaker when the Kubo number is small.
\end{abstract}

%In these cases inertial preferential sampling must matter, despite the fact that there is no small-scale fractal clustering of light particles in this inertia regime.

%\pacs{05.40.-a,47.55.Kf,47.27.eb}

% 05.40.-a Fluctuation phenomena, random processes, noise, and Brownian  motion
% 92.60.Mt Particles and aerosols
% 05.60.Cd Classical transport
% 45.50.Tn Collisions
% 47.27.-i Turbulent flows
% 05.40.Jc Brownian motion
% 47.55.Kf Particle-laden flows
% 47.27.eb Turbulence - Statistical theories and models
% 47.27.Gs Isotropic turbulence; homogeneous turbulence
% Find one PACS number of anisotropy, or gravity.

\maketitle

\section{Introduction}
Inertial particles moving in a turbulent flow do not simply trace the paths of fluid elements since particle inertia allows the particles to detach from the local fluid. This causes small-scale spatial clustering even
in incompressible turbulence~\citep{Max87,Dun05,Bec07,Vas08,Bra14}.
Moreover, {\it preferential sampling} of  regions with high vorticity/strain is  observed for particles that are lighter/heavier than the fluid.
As a result, the statistical properties of inertial heavy or light particles
 [for example water droplets in turbulent clouds~\citep{Sha03} or air bubbles in water~\citep{Maz03,Ren05}] in turbulence are highly non-trivial both at sub-viscous and larger scales.
An important  example is the case of  inertial particle accelerations \citep{Bec06c,Qur08,Gib10,Gib12,Vol11,Pra12}, see also
the review by \cite{TB2009}. At large inertia, singularities (caustics) occur
in the dynamics of heavy particles \citep{Fal02,Wil05}, giving rise
to large relative velocities between close  particles \citep{Sun97,Fal07c,Bec10,Gus11b,Gus13c},
and leading eventually to an enhancement of collision rates~\citep{Wil06,Bec14}.\\
%Remove or reduce?. During the last decade it has become clear that many aspects of the inertial dynamics of heavy
%particles in turbulence can be captured  in terms of statistical models that can be analysed rigorously \citep{Gus15}.
%Two essential simplifications make this possible: the turbulent velocity fluctuations are modeled by a statistical ensemble, and only Stokes force is taken into account in the equation of motion for the particle. The latter
%is a good approximation only for very small and very heavy particles.
The dynamics of light particles is important because of its relevance to  fundamental and applied questions. Light particles
can be used as small probes that  preferentially track high vorticity structures, highlighting statistical
 and topological properties of the underlying  fluid conditioned on those structures.  In the limit of high volume fractions
they might also have a complex feedback on the flow, including the case of reducing the turbulent drag \citep{Jac10,Ber05,Ber07}.
Compared to heavy particles the dynamics of light particles is more difficult to analyse because pressure-gradient
forces~\citep{Tch47} and  added-mass effects must be taken into account.
Apart from the fact that light particles tend to be drawn into vortices there is little theoretical analysis of their dynamics in turbulent flows.

This motivated us to formulate a {\it closure scheme} that allows us to compute inertial accelerations of both heavy and light particles in turbulence. We describe the scheme and  we test it by comparing its predictions to results of direct numerical simulations (DNS).
We consider a simple model for the particle dynamics, taking into account added-mass and pressure-gradient forces, but neglecting the effect of buoyancy on the acceleration as well as the Basset-Boussinesq history force, as in many previous DNS studies \citep{Bab00,Bec05,Cal08,Cal09,Vol08}. We comment on these questions in the conclusions.

Our scheme for the inertial particle accelerations approximates the particle paths by the Lagrangian fluid-element paths.
It is a closure for the particle equation of motion that neglects {\it inertial preferential sampling} such as heavy inertial particles being centrifuged out of long-lived vortical regions of the flow~\citep{Max87}, so that they preferentially sample straining regions, or light particles which by contrast are drawn into vortices \citep{Cal08,Cal08b}.
The latter is simply a consequence of the fact that
light particles are influenced by pressure gradients in the undisturbed flow.
Our approximation cannot be quantitatively correct when inertial preferential sampling is significant, but overall it yields a good qualitative description of
how particle accelerations depend on the particle density relative to that of the fluid, and upon the {\it Stokes number}, a dimensionless measure of particle inertia.
Moreover, and more interestingly, the closure also predicts highly non-trivial properties of the intermittent, non-Gaussian acceleration fluctuations, as the magnitude of inertial effects and the  Reynolds number of the turbulence change.  For example, the closure scheme predicts that the flatness of the acceleration  develops a  maximum or a minimum as a function of $\st$ for light or heavy particles. This is in good qualitative agreement with the measurements on the DNS data.
For heavy particles the closure  fails when  inertial preferential sampling affects inertial particle accelerations, i.e.  for Stokes numbers of order unity,
see Fig.~1{\bf b} in \citep{Bec06c}. In fact for heavy particles our closure is equivalent to the \lq filtering\rq{} mechanism
discussed by \cite{Bec06c}. For light particles, by contrast, comparisons to DNS results show
that  inertial preferential sampling effects are strong only at very  large Stokes numbers, $\st \sim 10$,
leading to an enlarged range of small and intermediate values of $\st$ where the closure works qualitatively well.

%Finally, by analysing a statistical model for the accelerations of heavy and light particles in turbulence we show
%that the above closure  can be obtained rigorously as a perturbation expansion within the model. The model contains
%an additional dimensionless parameter, the Kubo number. It characterises the degree to which flow structures persist in time.
%The perturbation expansion is developed  around the deterministic solution of the equation
%of motion, in the absence of turbulent fluctuations. In the limit of small Kubo numbers neither Lagrangian
%nor inertial preferential sampling matter, and the lowest order of the perturbation theory coincides with the closure scheme.
%It turns out that the perturbative expansion of the statistical model works well for the accelerations,
%even for intermediate Kubo numbers corresponding to flows that are persistent,
%yet less so than fully-developed turbulence. The statistical model is also in qualitative agreement with the results from  DNS of inertial particles in turbulence,
%the latter  showing slightly stronger preferential-sampling effects than the former at $\ku \sim 1$. At still larger Kubo numbers numerical simulations of the
%statistical model exhibit preferential-sampling effects that are in qualitative agreement with those observed in the DNS.

The remainder of this article is organized as follows. In the next section we formulate the problem,
introduce the equations of motion and qualitatively discuss inertial  preferential sampling.
 In Section \ref{sec:closure} we
describe our Lagrangian closure scheme for inertial particle accelerations.
%The statistical
%model and its perturbative analysis are discussed in Section \ref{sec:stat_model},
% where we also compare the analytical predictions with numerical results for the statistical model.
 Data from DNS of turbulent flows are analysed
in Section~\ref{sec:turbo} together with a detailed comparison to the predictions from the closure scheme.
In Section~\ref{sec:stat_model},
 we assess the potentialities of the closure on a data set obtained using a stochastic surrogate for the fluid velocity field.
Section~\ref{sec:conclusions}
contains the  conclusions.
%In an appendix we discuss the effect of Fax\'e{}n corrections that approximate the effect
%of particle size upon the acceleration, for particles smaller than the Kolmogorov length. We show how
%to take these corrections into account in the statistical model and demonstrate that the results are in good
%qualitative agreement with results from DNS.

\section{Formulation of the problem}
\label{sec:problem}
Many studies have considered the dynamics of heavy inertial particles, much denser than the carrying fluid.
When the particles are very heavy and at the same time very small (point particles)
the motion is simply determined by Stokes' drag.
The dynamics of light particles by contrast is also affected by pressure gradients of the unperturbed fluid,
and added-mass effects. Neglecting the effect of gravitational settling (and thus buoyancy) the equation of motion reads:
\begin{equation}
\begin{cases}
    \dot{\ve r}_t = {\ve v }_t\,,\\
    \dot{\ve v }_t = \beta  \D_t \ve u(\ve r_t,t) + \big(\ve u(\ve r_t,t)  - {\ve v}_t\big)\big/\taus\,,
    \label{mrfc}
\end{cases}
\end{equation}
where $ \ve r_t$ is the  particle position at time $t$, $ \ve v_t$, is the particle velocity,
$\ve u(\ve r_t, t)$ is the velocity field of the undisturbed fluid and
$\D_t\ve u = \partial_t \ve u + (\ve u\cdot\ve \nabla)\ve u$ is the Lagrangian derivative.
The dimensionless constant $\beta= 3\rho_{\rm f}/(\rho_{\rm f} + 2 \rho_{\rm p})$ accounts for the contrast between
particle density $\rho_{\rm p}$ and fluid density $\rho_{\rm f}$, while the Stokes number is defined as $\st = \taus/\tauK$
where the particle response time is $\taus = R^2 /(3 \nu \beta)$, $R$ is the particle radius,
$\nu$ the kinematic viscosity of the flow and $\tauK =\sqrt{\nu/\epsilon}$ the Kolmogorov time
defined in terms of the fluid energy dissipation, $\epsilon$. Many studies have employed
this model \citep{Bab00,Bec05,Cal08,Cal09,Vol08}.
The model takes into account added-mass effects but neglects buoyancy forces. It is an open question under which circumstances this is a quantitative model for the acceleration of small particles in turbulence.

The following analysis is based on Eqs.~(\ref{mrfc}).
It is important to first understand this case before  addressing more realistic situations (inclusion of buoyancy, finite size, collisions and feedback on the flow).
The sources of the difficulties are twofold. First,
even in the much simpler case of a non-turbulent  Eulerian velocity field $\ve u(\ve r,t)$
the particle dynamics is still complicated and often chaotic, simply because Eqs.~(\ref{mrfc}) are nonlinear.
Second, turbulence makes the problem even harder due to the existence of  substantial
spatial and temporal fluctuations. This results in chaotic Lagrangian dynamics of fluid elements.
Note that even though Lagrangian fluid elements sample space uniformly,
their instantaneous motion is in general correlated with structures in the underlying flow.
This implies that multi-time Lagrangian flow correlation functions evaluated along tracer trajectories
do not coincide with the underlying Eulerian correlation functions which are evaluated at fixed positions in space.
%We refer to this effect as {\it Lagrangian preferential sampling}.
Particles that are heavier or lighter than the fluid may detach from the flow if they have inertia.
This leads to the {\it inertial preferential sampling} mentioned in the introduction. As a result, the flow statistics
experienced by an inertial particle differs from that of a Lagrangian fluid element.

%Finally, we extend to the case of light particles
%the range of applicability of a previously developed  approach  based on the evolution of particles in a prescribed
%stochastic velocity  field with an  explicit expressions for all Eulerian fluid correlation functions \citep{Gus11a,Gus15}.
%In this statistical model Lagrangian preferential  effects can be calculated from the Eulerian statistics
%as a perturbation series in small values of the Kubo number.
%Retaining the lowest order in this expansion, neglecting the effects of Lagrangian preferential sampling, allows for explicit analytical solutions of the model.\\
\section{Lagrangian closure}
\label{sec:closure}
Let us notice that the dynamics determined by  equation (\ref{mrfc})
tends to the evolution of a tracer in both limits $\st \rightarrow 0$ and $\beta \rightarrow 1$. Indeed,
in the limit $\tau_s \rightarrow 0$, imposing  a finite Stokes drag leads to $\ve v_t = \ve u + O(\tau_s)$. For  $\beta = 1$, by contrast 
we may approximate the material derivative along fluid tracers trajectories with the derivative along the trajectory of a particle, $\D_t \ve u \sim \dot{\ve u}$, and  consistently check that  the evolution of (\ref{mrfc}) leads to an exponential relaxation of the evolution of the particle to the trajectory of the tracer.
%We can therefore be tempted to consider the closure as the lowest order
%term of a systematic expansion around tracer trajectories.
%We comment on this in the concluding section.
Hence, the idea is to approximate the effects of inertial forces on the particle trajectory
starting from the evolution of tracers. This approach cannot be exact.
It is for example known that vortices are preferentially sampled by inertial particles.
Nevertheless, it is important to understand and quantify how big the difference is as a function of the distance in the parameter phase space, $(\st, \beta)$, from the two lines $\st=0$ or $\beta=1$ where the closure must be exact (see. Fig. \ref{fig:phasespace}).
The starting point is to evaluate (\ref{mrfc}) along tracer trajectories:
\begin{equation}
\begin{cases}
    \dot{\ve r}^{({\rm L})}_t = \ve u(\ve r^{({\rm L})}_t,t)\,, \\
    \dot{\ve v_t} = \beta  \D_t   \ve u(\ve r^{({\rm L})}_t,t) + \big( \ve u(\ve r^{({\rm L})}_t,t)  - {\ve v_t}\big)\big/\taus\,,
    \label{mrfc2}
\end{cases}
\end{equation}
where $\ve r^{({\rm L})}_t$ denotes the Lagrangian trajectory of a tracer particle.
This approximation is a {\it closure} in the sense that the first equation in (\ref{mrfc2}) is independent of the second one,
and the second equation can be solved in terms of the Lagrangian velocity statistics $\ve u(\ve r^{({\rm L})}_t,t)$ of the underlying flow.
We solve (\ref{mrfc2}) for $\ve v_t$ (disregarding initial conditions here and below because these do not matter for the 
 steady-state statistics) to obtain
\begin{align}
\ve v_t&=\beta\ve u(\ve r_{t}^{(\rm L)},t)+\frac{1-\beta}{\taus}\int_0^t{\rm d}t_1e^{(t_1-t)/\taus}\ve u(\ve r_{t_1}^{(\rm L)},t_1)\,.
\eqnlab{v_implicit}
\end{align}
Using this expression, the particle acceleration follows from the second of Eqs.~(\ref{mrfc2})
\begin{align}
\ve a_t&=\beta\frac{{\rm D}\ve u}{{\rm D}t}(\ve r_{t}^{(\rm L)},t)+\frac{1-\beta}{\taus}\int_0^t{\rm d}t_1e^{(t_1-t)/\taus}\frac{{\rm D}\ve u}{{\rm D}t}(\ve r_{t_1}^{(\rm L)},t_1)\,.
\eqnlab{a_implicit}
\end{align}
Using \Eqnref{a_implicit} we express two-time acceleration statistics in terms of the two-point Lagrangian acceleration correlation function
\begin{equation}
C_{\rm L}(t) \equiv
\langle {\rm D}_t\ve u(\ve r_t^{(\rm L)},t)\cdot{\rm D}_t\ve u(\ve r_0^{(\rm L)},0)\rangle\,.
\end{equation}
In the steady-state limit we find
\begin{align}
\langle \ve a_t\cdot\ve a_0\rangle
&= \beta^2C_{\rm L}(t)+\frac{1-\beta^2}{\taus}\left[
\cosh\Big[\frac{t}{\taus}\right]\int_t^\infty{\rm d}t_1e^{-t_1/\taus}C_{\rm L}(t_1)\nonumber\\
&\hspace*{4cm}
+e^{-t/\taus}\int_0^t{\rm d}t_1 \cosh\left[\frac{t_1}{\taus}\right]C_{\rm L}(t_1)
\Big]\,.
\label{aCorr}
\end{align}
The acceleration variance is obtained by letting $t\to 0$ in Eq.~(\ref{aCorr})
\begin{equation}
\langle\ve a^2\rangle=\beta^2C_{\rm L}(0)+\frac{1-\beta^2}{\taus}\int_0^\infty{\rm d}t_1e^{-t_1/\taus}C_{\rm L}(t_1)\,.
\label{aVar}
\end{equation}
Similarly, the fourth moment of the particle acceleration is obtained as:
\begin{align}
\langle |\ve a|^4\rangle&
= \beta^4C_{\rm L}(0,0,0)
-4\frac{(\beta-1)\beta^3}{\taus}\int_0^\infty{\rm d}t_1e^{-t_1/\taus}C_{\rm L}(-t_1,0,0)\nn\\
& +6\frac{(\beta-1)^2\beta^2}{\taus^2}\int_0^\infty{\rm d}t_1\int_0^\infty{\rm d}t_2e^{-(t_1+t_2)/\taus}C_{\rm L}(-t_1,-t_2,0) \nn\\
& -\frac{(\beta-1)^3(3\beta+1)}{\taus^3}\int_0^\infty{\rm d}t_1\int_0^\infty{\rm d}t_2\int_0^\infty{\rm d}t_3e^{-(t_1+t_2+t_3)/\taus}C_{\rm L}(-t_1,-t_2,-t_3)\,.
\label{aQuad}
\end{align}
Here isotropy of the acceleration components $\langle a_ia_ja_ka_l\rangle=[\delta_{ij}\delta_{kl}+\delta_{ik}\delta_{jl}+\delta_{il}\delta_{jk}]\langle a_1^4\rangle/3$ was used to express $\langle |\ve a|^4\rangle$ in terms of the four-point Lagrangian correlation function
\begin{align}
C_{\rm L}(t_1,t_2,t_3)\equiv \frac{d(d+2)}{3}\langle {\rm D}_tu_1(\ve r_{t_1}^{(\rm L)},t_1){\rm D}_tu_1(\ve r_{t_2}^{(\rm L)},t_2){\rm D}_tu_1(\ve r_{t_3}^{(\rm L)},t_3){\rm D}_tu_1(\ve r_{0}^{(\rm L)},0)\rangle\,,
\end{align}
where ${\rm D}_tu_1$ is a component of the fluid acceleration and $d$ is the spatial dimension.
Eqs.~(\ref{aCorr}), (\ref{aVar}) and (\ref{aQuad}) express the fluctuations of inertial particle accelerations in terms of Lagrangian correlation functions
of the underlying flow, and predict how the inertial particle accelerations depend on $\beta$ and $\st$.
The integrals in Eqs.~(\ref{aVar}) and (\ref{aQuad}) are hard to evaluate numerically for very small and for very large values of $\st$.
When $\st$ is small, the exponential factors in the right-hand side of (\ref{aVar}) and (\ref{aQuad}) become singular and one needs a
 very high sampling frequency of the fluid acceleration along the particle trajectory to evaluate the integrals reliably.
When $\st$ is large, on the other hand, the integrals sum up large-time contributions with large fluctuations due to the
finite length of experimental and numerical  trajectories.
Before we proceed to a quantitative assessment of the model let us make a few general remarks about the range of applicability and accuracy of the approximation made. First, in Eqs. (\ref{aVar}) and (\ref{aQuad}) we need to evaluate the acceleration correlation function of the tracers for general values of $t$. This might be seen as a problem because the trajectories of tracers and of inertial particles must depart on a time scale of the order of the Lyapunov time. On the other hand, the acceleration correlation function of the tracer is known to decay on a time of the order of the Kolmogorov time, $\tau_K$, which is ten times smaller than the typical Lyapunov time of heavy particles \citep{Bec06} and stays close to zero in the inertial range of scales \citep{Falk12}. Therefore we do not see any problem in the evaluation of the integrals in the above equations. Second,  we are only interested in stationary properties of the inertial particle statistics and we always assume that the initial conditions for all particles are chosen from their stationary distribution functions. As a result no influence of the initial condition should appear in the closure.
The Lagrangian closure adopted here can also be applied to correlation functions of other observables of the particle, or of the flow, provided that the underlying Lagrangian correlation functions decay quickly enough.
%\kg{Note: I removed the discussion on the velocity correlation function because it did not seem to work very well when we checked (to slow decay).}
%one rewrites them in terms of the acceleration of the Lagrangian tracers. This is  always possible  by doing an  integration by parts in the equations for $\langle \ve v_t\cdot\ve v_0\rangle$.
Here,  we focus  on the acceleration statistics because of their highly non-Gaussian properties and 
high sensitivity to the parameters $\beta$ and $\st$. Finally, let us stress that our closure is fully based on Lagrangian properties and does require the use of any Eulerian correlation function, as opposed to closures based on the fast Eulerian approach by \cite{Fer01} or the  Lagrangian-Eulerian  closure
to predict two-particle distributions \citep{Zai03,Ali06,Der00,Pan10}.
To make further progress we need
to determine the Lagrangian correlation functions.
In Section \ref{sec:turbo} we present the most important new results, we determine the Lagrangian correlation functions by DNS of inertial particle dynamics in turbulence, substitute into Eqs.~(\ref{aCorr}), (\ref{aVar}) and (\ref{aQuad}) and assess the accuracy of these equations in predicting particle acceleration fluctuations and correlations.
%In Section \ref{sec:stat_model}
%we also show how to evaluate the correlations in a statistical model, where turbulent velocity $\ve u(\ve r,t)$ is less realistic and it is
%replaced by an appropriate stochastic function.

\section{Direct numerical simulations}
\label{sec:turbo}
\subsection{Simulation method}
\begin{figure}
\hspace*{-0.05in}
\includegraphics[width=13.5cm]{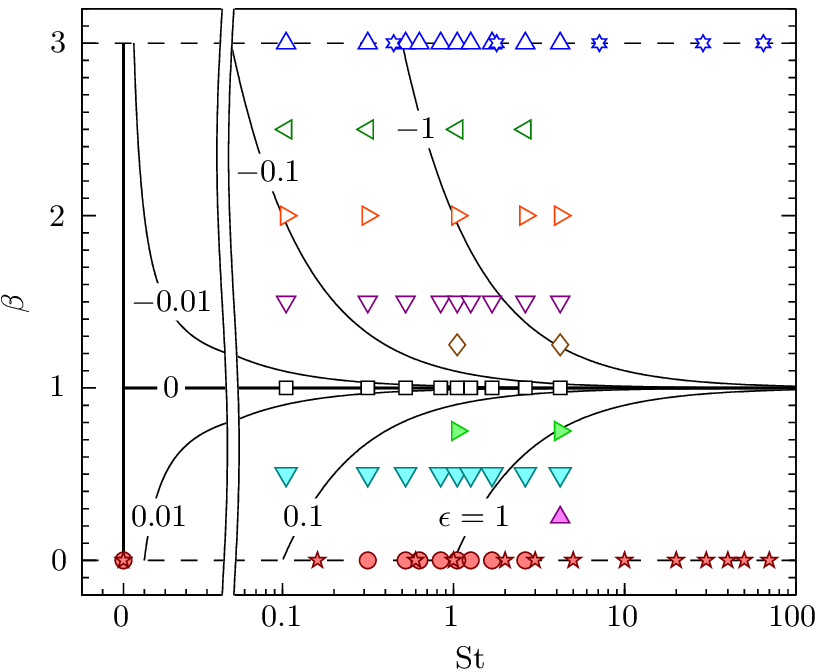}
\vspace{-0.2cm}
\caption{\small {\em (Online color).}
Parameter-space showing available DNS data sets with $\re_\lambda=185$ and $\re_\lambda=400$.
For values of $\st$ smaller than $0.05$ the $x$ axis is linear, while it is logarithmic for values of $\st$ larger than $0.05$.
For each data set with $\re_\lambda=185$ we have analysed a total of $130000$ trajectories of duration $6T_L$ and for each data set with $\re_\lambda=400$ we have analysed a total of $200000$ trajectories of duration $2.5T_L$.
Level curves $\st(1-\beta)=\epsilon$ for constants $\epsilon=\{-1,-0.1,-0.01,0,0.01,0.1,1\}$ are plotted as black lines.
Parameter families:
With $\re_\lambda=185$:
$\beta=0$ (red,$\circ$),
$\beta=0.25$ (magenta,$\vartriangle$),
$\beta=0.5$ (cyan,$\triangledown$),
$\beta=0.75$ (green,$\triangleright$),
$\beta=1$ (black,$\Box$),
$\beta=1.25$ (brown,$\Diamond$),
$\beta=1.5$ (purple,$\triangledown$),
$\beta=2$ (orange,$\triangleright$),
$\beta=2.5$ (dark green,$\triangleleft$),
$\beta=3$ (blue,$\vartriangle$).
With $\re_\lambda=400$:
$\beta=0$ (red,\marker{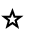}).
Additional data from~\cite{Cal09,Pra12}:
$\beta=3$ (blue,\marker{7}).
}
\label{fig:phasespace}
\end{figure}

We present here the analysis of data obtained
from  DNS of a homogeneous and isotropic turbulent flow seeded with point-like particles with different inertia and 
particle-fluid density
ratios  (see Fig.~\ref{fig:phasespace} for a summary of the $(\st,\beta)$ values available).
The data set was previously obtained by \citep{becJFM2010}. The flow obeys the  Navier-Stokes equations for
an incompressible velocity field ${\ve u}(\ve x,t)$:
\begin{equation}
  \partial_t\ve u + \ve u \cdot \nabla \ve u = -\nabla p +
  \nu\nabla^2\ve u +\ve f,\quad \nabla\cdot\ve u = 0\,.
  \label{eq:ns}
\end{equation}
The external forcing $\ve f$ is statistically homogeneous, stationary and isotropic,
injecting energy in the first low wavenumber shells, by keeping their spectral content constant~\citep{She}.
The viscosity $\nu$ is set such that the Kolmogorov
length scale $\eta\approx \delta x$, where $\delta x$ is the grid
spacing. The numerical domain is  $2\pi$-periodic
in the three directions. We use a fully dealiased
pseudospectral algorithm with second-order Adam-Bashforth
time stepping. For details see \citep{Bec06c,Bec06d}. Two series of DNS are analysed:
 Run I, with a numerical resolution of $512^3$ grid
points, and the Reynolds number at the Taylor scale $\re_\lambda
\approx 200$; Run II, with $2048^3$ resolution and $\re_\lambda \approx
400$. Details can be found in Table~\ref{table}.
\begin{table*}
  \centering

 \caption{\label{table} Eulerian parameters for the two runs analyzed
    in this Article: Run I and Run II in the text. $N$ is the number of grid
    points in each spatial direction; $\re_{\lambda}$ is the
    Taylor-scale Reynolds number; $\eta$ is the Kolmogorov dissipative
    scale; $\delta x=\mathcal{L}/N$ is the grid spacing, with
    $\mathcal{L}=2\pi$ denoting the physical size of the numerical
    domain; $\tauK=\left( \nu/\varepsilon \right)^{1/2}$ is the
    Kolmogorov dissipative time scale; $\varepsilon$ is the average
    rate of energy injection; $\nu$ is the kinematic viscosity;
    $t_{\mathrm{dump}}$ is the time interval between two successive
    dumps along particle trajectories; $\delta t$ is the time step;
    $T_L=L/U_0$ is the eddy turnover time at the integral scale $L=\pi$,
    and $U_0$ is the typical large-scale velocity.}
\mbox{}\\
  \begin{tabular}{ccccccccccc}
    \hline\\[-7pt]
    & $N$ & $\re_{\lambda}$ & $\eta$ & $\delta x$ & $\varepsilon$ &
    $\nu$ & $\tauK$ & $t_{\mathrm{dump}}$ & $\delta t$ & $T_L$
    \\[+2pt]\hline\\[-7pt] Run I& 512 & 185 & 0.01 & 0.012 & 0.9 & 0.002 &
    0.047 & 0.004 & 0.0004 & 2.2 \\ Run II& 2048 & 400 & 0.0026 & 0.003 &
    0.88 & 0.00035 & 0.02 & 0.00115 & 0.000115 & 2.2\\[+2pt]\hline
  \end{tabular}
\end{table*}

\subsection{Comparison for acceleration variance and flatness against DNS}
\label{sec:dns_results}
%\begin{figure}
%\includegraphics[width=13.75cm]{figs/Fig4.pdf}
%\vspace{-0.2cm}
%\caption{\small {\em (Online colour).} \lb{we need to discuss all symbols!!!!}
%Acceleration variance and flatness for DNS data.
%Lines show the closure scheme prediction for ({\bf a}) the acceleration variance, Eq.~(\ref{aVar}), and ({\bf b}) the flatness obtained by Eqs.~(\ref{aVar}) and (\ref{aQuad})
%using numerically evaluated Lagrangian acceleration correlation functions with $\re_\lambda=185$ (solid lines) and $\re_\lambda=400$ (dashed red lines).
%Thin dashed black line in panel ({\bf b}) shows the limit of Gaussian distributed acceleration components, \Eqnref{GaussianFlatness}.
%}
%\figlab{acc_var_DNS}
%\end{figure}
In Fig.~\ref{fig:accvarDNSotherbetas}{\bf a} we show the comparison of the acceleration variance, $\langle a^2 \rangle$,
 to DNS data as a function of $\st$
for different values of  $\beta$.
Consider first heavy particles ($ \beta <1 $).  It is clear that the closure (\ref{mrfc2}) captures the general trend and it
becomes better and better for larger Stokes numbers. Similarly, this approximation must become exact as $\st \to 0$, but
 the DNS data set does not contain values of $\st$ small enough to reach this limit.
At intermediate Stokes numbers the  Lagrangian closure described in  Section \ref{sec:closure} does not
match the DNS results.  This mismatch for Stokes numbers between $0.1$ and $1$ is certainly
 due to preferential sampling, as already remarked  by \cite{Bec06c}. Yet, the closure predicts a small Reynolds dependency (compare solid and dashed red lines for $\beta=0$)
inherited  by the dependency of the fluid tracers. Such  a small variation is not detectable within
 the accuracy of our numerical data.
Now consider light particles ($\beta >1$). Fig. ~\ref{fig:accvarDNSotherbetas}{\bf a} shows  that also in this case the agreement between the
 DNS data and the closure scheme is good,  except around $\st \sim 10$ where the closure underestimates the DNS data.
\begin{figure}
\hspace*{-0.05in}
\includegraphics[width=13.5cm]{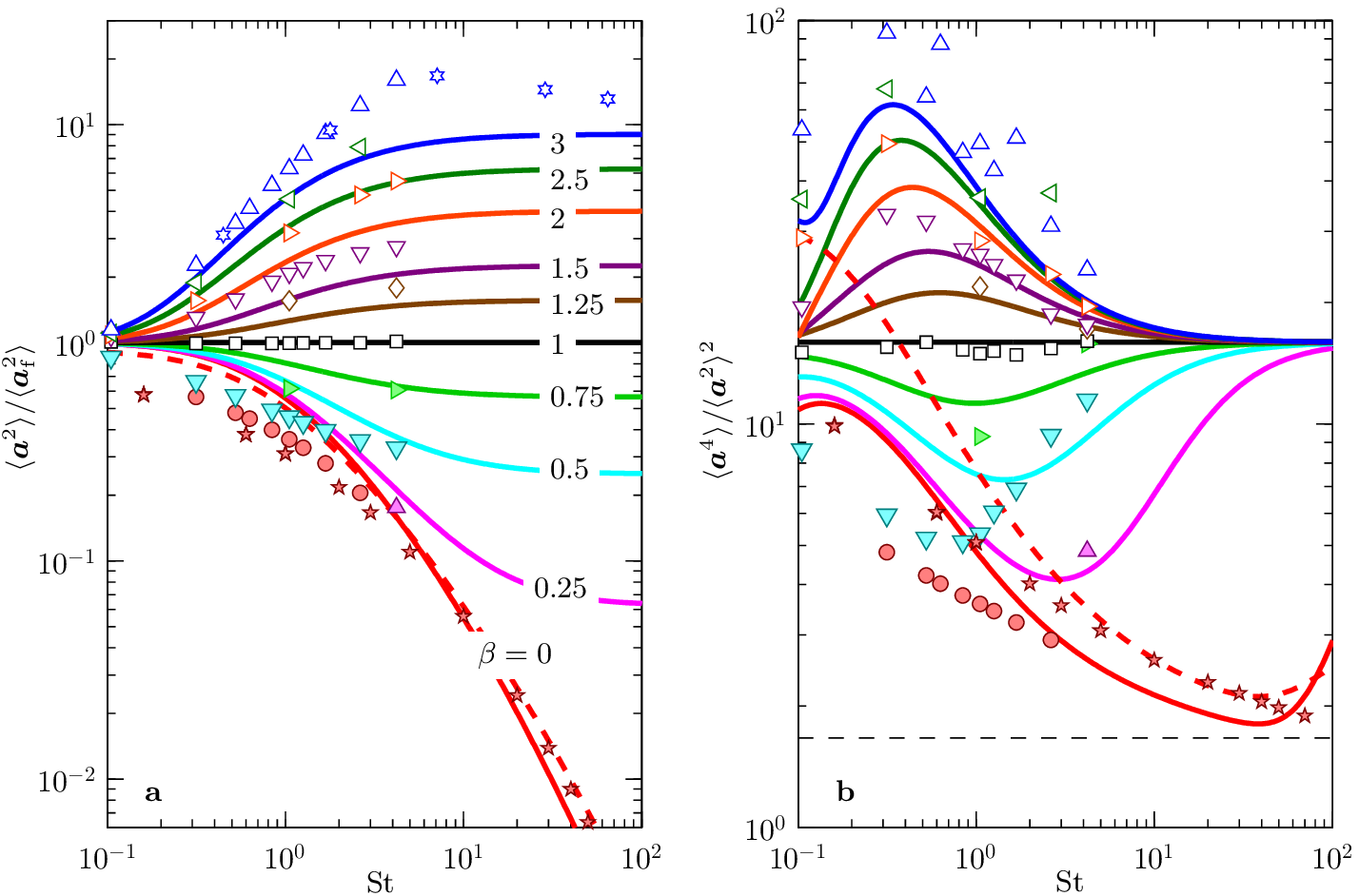}
\vspace{-0.2cm}
\caption{\small {\em (Online colour).}
Acceleration variance ({\bf a}) and flatness ({\bf b}) for DNS data at changing $\beta$, $\st$ and Re.
{\bf a}: points correspond to the DNS data (same symbols of Fig.\ref{fig:phasespace}). Solid lines (labeled with their corresponding value of $\beta$) show the closure scheme prediction for the acceleration variance of light and heavy particles, Eq.~(\ref{aVar}), normalized with the fluid variance, for all data from  RUN I;
dashed line corresponds to the closure for RUN II.
{\bf b}: the flatness measured on  the DNS data and the one
predicted by the closure scheme, Eqs.~(\ref{aQuad}), for heavy and light particles.
Thin dashed black line shows the limit of normal distributed acceleration components.
Additional data for $\beta=3$ (blue,\protect\includegraphics[width=2mm,clip]{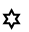}) is omitted in panel {\bf b} because the flatness was not evaluated in Refs.~\cite{Cal09,Pra12}.
}
 \figlab{accvarDNSotherbetas}
\end{figure}
%\begin{equation}
% \langle  a^2 \rangle > \beta^2 \langle \ve a_{\rm f}^2\rangle\,.
%\end{equation}
In this case inertial preferential sampling must be  important. The reason
is that light particles are drawn into vortex filaments where they experience high accelerations.
Nevertheless, inertial preferential sampling must  become irrelevant in the limit $\st\to\infty$ as shown by the trend for very large $\st$ in the same figure.
%Let us also notice  that \faxen corrections
%reduce the acceleration variance of light particles at large Stokes numbers \citep{Cal09,Pra12} because the particles
%effectively average over small-scale fluid-velocity fluctuations. This effect is computed in the statistical model in Appendix~A.
It is interesting to
 remark that  the closure scheme  (solid line) becomes better and better the closer $\beta$ is to unity,  and/or the
smaller the Stokes number is, suggesting the possibility to develop a systematic perturbative expansion in the small parameter
$\epsilon = \st (1-\beta)$.
 Furthermore it is important to note that while the acceleration variance increases  monotonically as both $\beta$ and $\st$ increase,
the flatness of the acceleration, defined as
$$F_{\st,\beta} = \langle |\ve a|^4\rangle/\langle \ve a^2\rangle^2,$$ has a non-monotonic dependency on $\st$.
%\kg{I rewrote this as I though the previous discussion on intermittency to be confusing:}
In  Fig.~\ref{fig:accvarDNSotherbetas}{\bf b} we show that light particles have a maximum in their flatness at $\st \sim 0.5$
 and heavy particles have minimum flatness for $\st>1$ (except for the case of very heavy particles with $\beta=0$).
%In panel {\bf b} of Fig.~\ref{fig:accvarDNSotherbetas} we show that there is a clear maximum of intermittency for $\st \sim 0.5$ (except for the case of very heavy particles with $\beta=0$).
Importantly, our closure approximation predicts these extrema qualitatively,
indicating that the  non-Gaussian tails observed numerically and experimentally \citep{Vol08} in the acceleration probability distribution function  of bubbles
 are not only due to  inertial preferential sampling, which is neglected by the closure.
Let us also note that the non-monotonicity shown by the model for $\st \sim 0.1$ and $\beta=0$, $0.25$ and $3$ are artefacts due to insufficient accuracy in the numerical evaluation of the integral in Eq.~(\ref{aQuad}).
In conclusion, remarkably, the closure approximation is in reasonable agreement with the DNS data, with the exception of those values of Stokes number where preferential
 sampling is important. This is an important result, supporting the idea that many key properties, including deviations from Gaussian statistics,
 of the acceleration statistics of inertial particles are due to the kinematic structure of the equation of
motion together with the non-trivial Lagrangian properties of the
fluid tracers evolution, and not only due to inertial preferential sampling.
%Compared to the acceleration variance the flatness is somewhat more influenced by preferential sampling, at least for heavy particles.
%\lb{We stress again
%that  the numerical instabilities of the solid lines shown in \Figref{acc_var_DNS}{\bf b} for extreme values of $\st$ are due to the difficulties to
%evaluate numerically the integrals  in Eqs.~(\ref{aVar}) and (\ref{aQuad}).}

\begin{figure}
\hspace*{-0.05in}
\includegraphics[width=13.5cm]{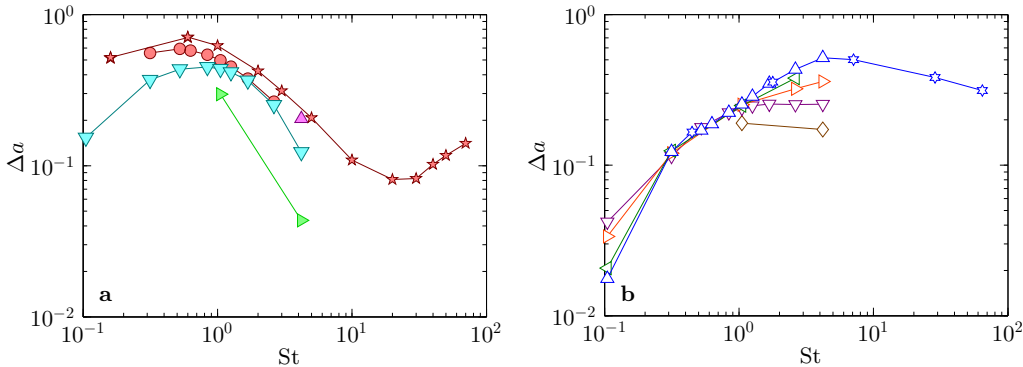}
\vspace{-0.2cm}
\caption{\small {\em (Online colour).}
Relative error $\Delta a^2$ \eqnref{relerr} between the DNS data and the closure scheme prediction for heavy particles ({\bf a}) and light particles ({\bf b}). Markers according to \Figref{phasespace}. Lines are drawn as a guide to the  eye.}
\figlab{acc_error_DNS}
\end{figure}
In order to quantify the accuracy of the closure described in \Secref{closure} we plot in Fig.~\ref{fig:acc_error_DNS} the relative error in the prediction of the acceleration variance
for both light and heavy particles:
\begin{equation}
\Delta a^2 = \left|1 - \frac{\langle \ve a^2 \rangle}{\langle \ve a^2 \rangle_{\rm DNS}}\right|\,,
\eqnlab{relerr}
\end{equation}
where with $\langle \ve a^2 \rangle$ we refer to the expression (\ref{aVar}) and with
$ \langle \ve a^2 \rangle_{\rm DNS}$ to the numerical value obtained from the DNS results. As one can see the approximation is never very bad, with a maximum discrepancy of the order of $30-40\%$ at those values where inertial preferential sampling is important, i.e. for $\st \sim O(1)$ for heavy particles and
for $\st \sim O(10)$ for light particles.

%\begin{figure}
%\hspace*{-0.05in}
%\includegraphics[width=13.5cm]{figs/Fig8.pdf}
%\vspace{-0.2cm}
%\caption{\small {\em (Online colour).}
%Comparison between the acceleration correlation function from DNS data and the prediction from the closure scheme for $\st=0.5$ ({\bf a}) and $2.6$ ({\bf b}) and for three different $\beta$: $\beta = 0$ (red,$\circ$), $\beta = 1$ (black,$\Box$), and $\beta = 3$ (blue,$\vartriangle$).
%}
%\figlab{acc_corr_DNS}
%\end{figure}

\section{Statistical Eulerian velocity model}
\label{sec:stat_model}
Mathematical analysis and numerical studies of the particle dynamics become easier when the turbulent
fluctuations of $\ve u(\ve r,t)$ are approximated by a stochastic process.
Following \cite{Gus15} we use a smooth, homogeneous and isotropic
Gaussian random velocity field with root-mean-squared speed $u_0$ and typical length and
time scales $\eta$ and $\tau$.
The model is characterized by a dimensionless number, the Kubo number, $\ku= \tau/(\eta/u_0)$,
that measures the degree of persistence of flow structures in time. Very small Kubo numbers correspond
to a rapidly fluctuating fluid velocity field. In this limit the closure approximation
described in \Secref{closure} is exact, inertial preferential sampling is negligible and
the Lagrangian correlation functions of tracer particles are well approximated by the Eulerian correlation functions.
In this limit it is also possible to perform a systematic perturbative expansion \citep{Gus15}. In this paper
we are  interested in comparing the validity of the Lagrangian closure for the statistical model at $\ku \sim O(1)$, where no analytical results can be obtained and to further compare them with the DNS results shown in \Secref{turbo}. The motivation  is the following. The statistical model has no 'internal intermittency', i.e. there is no Reynolds number dependency on the acceleration statistics (there is not even  the meaning of a Reynolds number). Nevertheless, once the Gaussian Eulerian velocity field is prescribed, we can calculate the acceleration probability density function of the fluid tracers. It turns out that this is not Gaussian and that it depends on the Kubo number,  due to the effect of the quadratic advection term, $\ve a_{\rm f}=\partial_t\ve u+\ku[\ve\nabla\ve u\T]\ve u$. As a result, we expect that many of the properties shown by the acceleration distribution of
inertial particles evolved in real turbulent flows are shared by particles evolved in a Gaussian random flow.
Finally, a comparison between DNS data and the statistical model will allow us to assess further the importance of internal intermittency.
\subsection{Construction of the random velocity field}
For simplicity we discuss only the two-dimensional case. Generalization to three dimensions is straightforward.
The velocity field is given in terms of the streamfunction: $\ve u(\ve r,t)=\ve\nabla\psi(\ve r,t)\wedge\hat{\ve e}_3\,$, which is defined
as a superposition of Fourier modes with a Gaussian cutoff,
\begin{equation}
\label{eq:psi}
\psi(\ve x,t)=\frac{\eta^2u_0}{\sqrt{\pi}L}\sum_{\sve k}a_{\sve k}(t)e^{{\rm i}\sve x\cdot\sve k-k^2\eta^2/4}\,.
\end{equation}
Here the system size $L$ is put to $10\eta$, $k_i=2\pi n_i/L$ and $n_i$ are integers with an upper cutoff $|n_i|\le 2L/\eta$ because higher-order Fourier modes are negligible.
The resulting spatial correlation function of $\psi$ is Gaussian, if $L\gg\eta$ we have:
$\langle\psi(\ve x,0)\psi(\ve 0,0)\rangle=(u_0^2\eta^2/2)\,e^{-x^2/(2\eta^2)}.$
The random coefficients $a_{\sve k}(t)$ in $\psi$ are drawn from random Gaussian distributions with zero means, smoothly
correlated in time. To do that we used  an Ornstein-Uhlenbeck process  convolved  with a Gaussian kernel of the form
$
w(t)\equiv \exp[-t^2/(2t_0^2)]/(t_0\sqrt{2\pi})\,$ to have a smooth correlation function also for the acceleration.
The parameter $t_0$ must be small, $t_0\ll \tau$, in order for the flow field to
decorrelate at long times in a similar fashion as in fully developed turbulence.
The
Eulerian autocorrelation function of ${\ve u}$ is:
\begin{equation}
    \langle {\ve u}(\ve x_0,t)\cdot {\ve u}(\ve x_0,0)\rangle =
    \frac{u_0^2}{2\sqrt{\pi}}e^{-t^2/(4t_0^2)} \left(\E\left[\frac{t_0}{\tau} + \frac{t}{2t_0}\right]
    \right.
    + \left. \E\left[\frac{t_0}{\tau}-\frac{t}{2t_0}\right]\right)\,,
    \eqnlab{uCorrSmoothMain}
\end{equation}
where $\E(x) = \sqrt{\pi}\exp(x^2)\erfc(x)$.
Note that the flow field is homogeneous in space and also in time and
  $\langle{\ve u}(\ve x_0,t)\cdot{\ve u}(\ve x_0,0)\rangle$
is a function of $|t|$ only. For $ |t| \ll t_0$ this correlation
function is Gaussian and for $ |t| \gg t_0$ the correlation
function is exponential:
$\langle{\ve u}(\ve x_0,t)\cdot{\ve u}(\ve x_0,0)\rangle\sim e^{-|t|/\tau}$.
\subsection{Results for the random velocity model}
\label{sec:res_stat_model}
\begin{figure}
\hspace*{-0.05in}
\includegraphics[width=13.7cm]{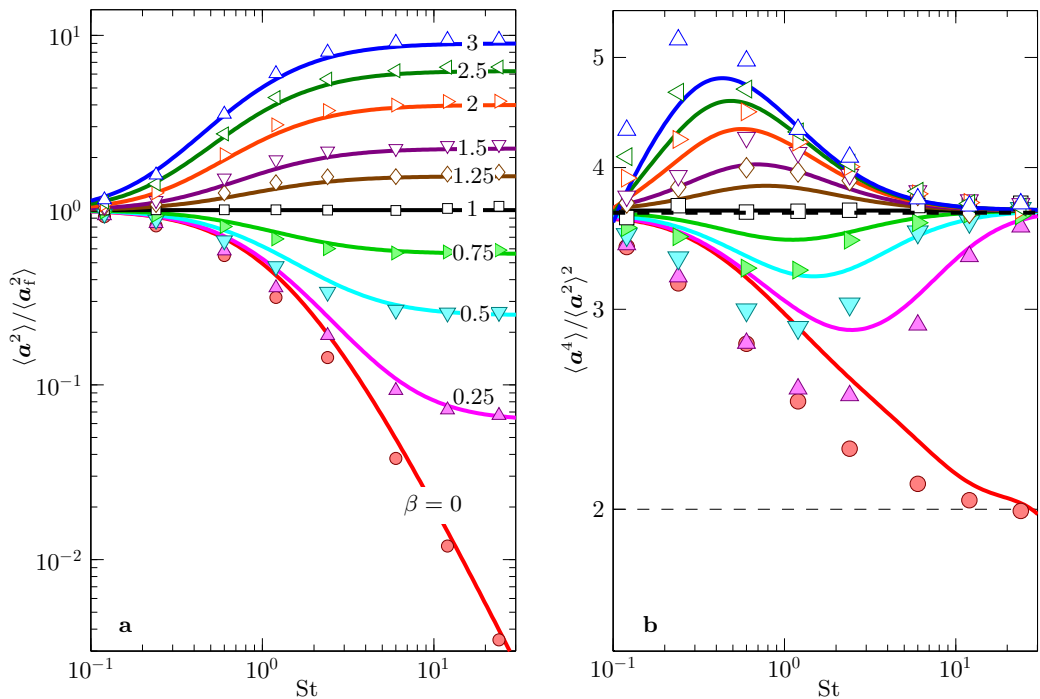}
\vspace{-0.2cm}
\caption{
\small {\em (Online colour).}
Same as in Fig. \ref{fig:accvarDNSotherbetas} but for the statistical model given by Eq. (\ref{eq:psi}) at $\ku=5$.
%Acceleration variance normalized by the fluid acceleration variance ({\bf a}) and acceleration flatness ({\bf b}) against $\st$ for the statistical model.
%Symbols show results from numerical simulations of particles following Eq.~(\ref{mrfc}) with $\ku=1$ (medium-size markers) and $\ku=5$ (large-size markers).
%Solid lines show theory in the limit of small Kubo numbers for the acceleration variance in panel {\bf a}, \Eqnref{a2theo}, and flatness in panel {\bf b} (limit of normal distributed acceleration components), \Eqnref{GaussianFlatness}. Black-dashed line in panel {\bf a} shows asymptote \eqnref{a2LargeSt} for large values of $\st$, $\langle\ve a^2\rangle/\langle\ve a_{\rm f}^2\rangle\sim\beta^2$.
%Black-dashed lines in panel {\bf b} correspond to the theory for the fluid flatness, \Eqnref{FluidFlatness}, with $\ku=1$ (lower line) and $\ku=5$ (upper line).
%Parameters: $d=2$, $\tau=20t_0$, $t_0=3\tauK/4$, $\beta = 0$ (red,$\circ$), $\beta = 1$ (black,$\Box$), and $\beta = 3$ (blue,$\vartriangle$).
}
\figlab{acc_var_stat}
\end{figure}
We consider first the acceleration variance. Simulation results for the statistical model are compared with the Lagrangian closure (\ref{aVar}-\ref{aQuad}) in Fig.~\ref{fig:acc_var_stat} by changing both $\st, \beta$  for $\ku=5$  (panel {\bf a}).
We observe a good agreement between the Lagrangian closure and the numerical simulations, comparable to what was observed for the
DNS data.
In \Figref{acc_var_stat}{\bf b} we show the results for the flatness. It is important to stress two facts. Also here, both  heavy and light particles depart from the corresponding fluid value with a qualitative trend similar to that observed for the DNS case in the previous section. Also for the random velocity field, the Lagrangian closure works qualitatively well. The departure from the numerical data is a signature of the corresponding importance of preferential sampling at those Stokes numbers.
Let us notice nevertheless an important difference with respect to the DNS  data. Here the absolute values of the flatness are much smaller, due to the absence of internal intermittency. In the stochastic signal the acceleration of the fluid is non-Gaussian only because of
kinematic effects. In real flows, the acceleration is  more intense and  more fluctuating because of the vortex stretching mechanism and of the turbulent energy cascade.

\section{Conclusions}
\label{sec:conclusions}
In this paper we have analysed a Lagrangian closure  describing fluctuations and correlations
of inertial particle accelerations in turbulent flows in the diluted regime (one-way coupling), i.e. neglecting particle-particle collisions and feedback on the flow. In this way, we have a model that is able to predict some properties of the
acceleration statistics of inertial particles for a large range of values of $\beta$ and $\st$ out of one single measurement based on fluid tracers only. We have compared the predictions of the closure
to DNS of heavy and light inertial particles in turbulence.  To summarize our results, the closure predictions are in overall
good qualitative agreement  with the results of DNS of particles.
The closure neglects inertial preferential sampling, i.e. the tendency of light/heavy particles to be centrifuged in  or out of vortex structures. Hence, the good agreement with the DNS data indicates
 that inertial preferential sampling has in general only a partial effect on inertial particle accelerations.
The main trends are essentially kinematic, a consequence of the form of the equation of motion, as also shown by
the results obtained using a stochastic surrogate for the flow velocity.
A closer inspection shows that there are important differences between the Lagrangian closure scheme and the DNS,
revealing where inertial preferential sampling is important. The effect is larger for light particles
at large Stokes numbers ($\st \sim 10$ in our DNS), and is a consequence of the fact that light bubbles
are drawn into intense vortex tubes. We mention that there is no small-scale fractal clustering for these values of $\st$, i.e. particles are distributed on a three-dimensional set at scales much smaller than the Kolmogorov length.
Finally, non-trivial non-monotonic behaviours of the flatness for both light and heavy particles as a function of $\st$
are predicted by the closure scheme and confirmed by the DNS results, including the fact that light particles are always more intermittent than the fluid tracers and the opposite holds for heavy particles, as shown by the fact that the flatness for the former is always larger than the one of fluid tracers and vice versa for the latter.
The Lagrangian closure scheme must become exact when $\st \to 0$ or $\beta \to 1$, it should therefore be possible to see it
 as a perturbative expansion around Lagrangian tracers and proportional to a small parameter $\epsilon= \st(1-\beta)$,
at least for quantities that depend on Lagrangian correlation functions  decaying on a time scale of the order of the Kolmogorov time.
%with a possible correction in the functional dependence to  $f(\st)(1-\beta)$ for  large Stokes numbers.
In this case, one could try to develop an intermediate asymptotic where for small enough time the difference between the two trajectories remains small and then improve the zeroth-order approximation here presented by considering also corrections induced by the velocity gradients around the Lagrangian tracers:
\begin{equation}
 u^i (\ve r_t,t) \sim  u^i (\ve r^{({\rm L})}_t,t) + \partial_i u^j(\ve r^{({\rm L})}_t,t) \delta  r^j_t + \cdots\,,
\end{equation}
where $\delta\ve r=\ve r_t-\ve r_t^{(\rm L)}$.
Work in this direction is in progress.  Finally, we have also investigated the validity of the Lagrangian closure using a stochastic Gaussian surrogate for the advecting fluid velocity field. In such a case, the $\ku$ number is another free parameter that can be tuned to increase/decrease the effects of inertial preferential sampling (effects vanish as $\ku$ approaches zero). We have shown that for large Kubo numbers, corresponding to the long-lived structures in turbulent flows, the closure theory works as well as for the DNS data, even though the data for the statistical model have a much smaller flatness.\\
Let us add some  remarks about the generality and the  limitations of the approach proposed.
First, there are no theoretical difficulties in incorporating  buoyancy, Fax\'{e}n corrections and other forces in the closure scheme as long as the dynamics
 can be described by a point-particle approach. We refrained from presenting here the results because of lack of DNS data to compare with. On the other hand,
 it is known that the equations (\ref{mrfc}) are not valid for all values in the ($\beta,\st$) parameter space. Indeed,
the two requirements that the Reynolds number based on the particle slip velocity is small: $Re_p = |u-v|R/\nu < O(1)$ and that the particle size is smaller
than  the Kolmogorov scale $R/\eta < O(1)$  lead to the condition that $\st < O(1)$ if $\beta > 1$. So the prediction of the model in the limit of large Stokes numbers for light
particles cannot be taken on a quantitative basis.  We stress nevertheless that the most interesting property highlighted by our approach, i.e. the existence of a non-monotonic behaviour for the flatness of the acceleration of light and heavy particles, develops at values of $\st$ where the model equations are still valid. It is difficult to precisely assess the value of Stokes where the approximation breaks down. For instance, recently  it was found \citep{Lohse2015}  that the acceleration variance of light particles with a size up to $R \sim 10\eta $ follow quite closely the point-like approximation (\ref{mrfc}). For even larger particle sizes a wake-driven dynamics  becomes dominant. For such a range of particle parameters no theoretical models for the equations of motion are known.
For instance, recently it was found (Mathai et al. 2015) that the acceleration variance of light particles with a size up to R ∼ 10η follow quite closely the approximate dynamics (2.1). 

{\em Acknowledgments.}
This work was supported Vetenskapsr\aa{}det and by the grant {\em Bottlenecks for particle growth in turbulent aerosols} from the Knut and Alice Wallenberg Foundation, Dnr. KAW 2014.0048. The research leading to these results has received funding from the European Union's Seventh Framework Programme (FP7/2007-2013) under grant agreement No 339032.

%\bibliographystyle{jfm}
%\bibliography{biblio,settling}

\end{document}